\documentclass[a4paper,12pt]{article}
\usepackage{amsbsy}
\usepackage{amsmath}

\begin{document}

\title{Invariance group important for the interpretation of Bose-Einstein correlations}
\author{K. Zalewski\\
M.Smoluchowski Institute of Physics Jagellonian University, Cracow\\
and\\  Institute of Nuclear Physics of the Polish Academy of Sciences }
\maketitle

\begin{abstract}

A group of transformations changing the phases of the elements of the
single-particle density matrix, but leaving unchanged the predictions for
identical particles concerning the momentum distributions, momentum
correlations etc., is identified. Its implications for the determinations of
the interaction regions from studies of Bose-Einstein correlations are
discussed.
\end{abstract}

\noindent PACS 25.75.Gz, 13.65.+i \\Bose-Einstein correlations.\vspace{1cm}

\section{Introduction}
Bose-Einstein correlations are helpful when trying to derive properties of the
interaction regions from the measured momentum distributions. In this report we
point out an ambiguity inherent in such derivations \cite{BIZ}. There are many
ways from the data to the inferred properties of the interaction regions (cf.
e.g. \cite{WIH} and references given there). One can use density matrices,
Wigner functions, emission functions, distances between the pairs of points
where the identical particles are produced etc. The ambiguity seems to be
common to all of them.

We will not discuss here the evolution of the density matrix during the
freeze-out period. Formally, one can introduce the assumption that all the
hadrons have been produced instantaneously and simultaneously at some time
which we may choose as t=0. Then what one measures is the density matrix (in
the interaction representation) at freeze-out.

As is well known, the diagonal elements of the $k$-particle density matrix in
the momentum representation give, and are unambiguously given by, the $k$
particle momentum distribution. On the other hand, in most models these
elements can be expressed as symmetrized products of the single particle
density matrix elements \cite{KAR}:

\begin{equation}\label{karczm}
  \rho(\textbf{p}_1,\ldots,\textbf{p}_k;\textbf{p}_1,\ldots,\textbf{p}_k) =
  \sum_P\prod_{j=1}^k \rho_1(\textbf{p}_j;\textbf{p}_{Pj}),
\end{equation}
where the summation is over all the $k!$ permutations of the momenta
$\textbf{p}_1,\ldots,\textbf{p}_k$. Thus, all the momentum distributions are
unambiguously determined when the single particle density matrix
$\rho_1(\textbf{p}_1;\textbf{p}_2)$ is known.

Our main observation \cite{BIZ} is that the converse is not true. Given the
momentum distributions for all the sets of $k = 1,2,\ldots$ particles, it is
not possible to find unambiguously the matrix $\rho_1$. This trivial
observation will be seen to have very non trivial consequences. Since the
matrix $\rho_1$ is further used the derive conclusions about the interaction
region, the ambiguity affects our capacity for making unambiguous statements
about such regions.

\section{Invariance group}

Consider the transformation

\begin{equation}\label{invgro}
  \rho_1(\textbf{p};\textbf{p'}) \rightarrow \rho_{1\alpha}(\textbf{p};\textbf{p'}) \equiv
  e^{i\alpha(\textbf{p})}\rho_1(\textbf{p};\textbf{p}')e^{-i\alpha(\textbf{p'})},
\end{equation}
where $\alpha$ is any real-valued function of momentum. According to formula
(\ref{karczm}) for $k=1$, $\rho_1$ is a single particle density matrix.
Therefore, it must be hermitian and have trace one. Also $\rho_{1\alpha}$, as
seen from its definition, is hermitian and has trace one. Consequently, it can
be substituted for $\rho_1$ on the right-hand side of formula (\ref{karczm}).
This introduces on the right-hand side, for every $p_i$ a term
$e^{i\alpha(\textbf{p}_i)}$ and a term $e^{-i\alpha(\textbf{p}_i)}$ which
cancels it. Thus, the diagonal matrix element on the left-hand side does not
change. Experimentally, the substitution of $\rho_{1\alpha}$ for $\rho_1$ is
invisible. The transformations from $\rho_1$ to $\rho_{1\alpha}$ form a local
(in momentum space) U(1) invariance group.

There are several quantities related to the single particle density matrix in
the momentum representation and yielding information about the interaction
region. In order to get the space distribution of the sources one can use the
diagonal elements of the single particle density matrix in the coordinate
representation

\begin{equation}\label{}
  \tilde{\rho}_1(\textbf{x};\textbf{x}) = \int\!\!d^3K\frac{d^3q}{(2\pi)^3}\;e^{i\textbf{q}\cdot \textbf{x}}\rho_1(\textbf{p};\textbf{p'}),
\end{equation}
where

\begin{equation}\label{}
  \textbf{K} = \frac{1}{2}(\textbf{p} + \textbf{p'});\qquad \textbf{q} = \textbf{p} - \textbf{p'}.
\end{equation}
In order to get an approximate phase-space distribution one can use the Wigner
function

\begin{equation}\label{}
  W_1(\textbf{K},\textbf{X}) = \int\!\!\frac{d^3q}{(2\pi)^3} e^{i\textbf{q}\cdot \textbf{X}}\rho_1(\textbf{p};\textbf{p'}),
\end{equation}
or the emission function

\begin{equation}\label{}
  \rho_1(\textbf{p};\textbf{p'}) = \int\!\!d^4X S(K,X)e^{iqX}.
\end{equation}
In the last two formulae $X = \frac{1}{2}(x + x')$ and the approximation
consists in interpreting $K$ and $X$ as the energy-momentum and space-time
position of the particle. This approximation is not always good, but in a
well-defined sense \cite{HIL} this is the best one can have without
contradicting the principles of quantum mechanics. In the last formula the
four-vector $K$ has the same value on both sides of the equality. Since the
momenta $\textbf{p}$ and $\textbf{p'}$ are on mass shell, for any $\textbf{q}$
the value of $q_0$ is fixed by the condition $Kq=0$. Thus, it is not possible
to invert the Fourier transform and to express the emission function $S$ in
terms of the density matrix $\rho_1$. In fact, there is an infinity of
different emission functions, corresponding to various off mass shell
continuations of a given on mass shell function  $\rho_1$. We conclude that the
ambiguities when using the emission function formalism are more severe than
when using Wigner functions.

The point is that the transition from $\rho_1$ to $\rho_{1\alpha}$, which has
no effect on the momentum distributions, changes the functions
$\tilde{\rho}_1$, $W_1$ and $S$ and consequently changes the conclusions
concerning the interaction region. In the following section we will illustrate
this fact by some examples.
\newpage
\section{Examples}.

The class of transformations (\ref{invgro}) is very rich. We will just discuss
three simple examples. Consider

\begin{equation}\label{}
  \alpha(\textbf{p}) = \textbf{b}\cdot \textbf{p} \qquad \Rightarrow\qquad \alpha(\textbf{p}) - \alpha(\textbf{p'}) =
  \textbf{b}\cdot \textbf{q},
\end{equation}
where $\textbf{b}$ is any vector. This gives

\begin{equation}\label{}
  \tilde{\rho}_1(\textbf{x};\textbf{x}) = \int\!\!dK\frac{dq}{(2\pi)^3}\;e^{i\textbf{q}\cdot(\textbf{x} +
  \textbf{b})}\rho_1(\textbf{p};\textbf{p'}).
\end{equation}
The interaction region gets shifted by $\textbf{b}$. Similarly, replacing
$\textbf{b}$ and $\textbf{p}$ by four-vectors and using the emission function
one can generate an arbitrary shift in space-time. This result is of little
interest. It is obvious that the momentum distributions do not depend on where
and when the experiment was done.

As our second example consider

\begin{equation}\label{}
  \alpha(\textbf{p}) = \frac{1}{2}c\textbf{p}^2 \qquad \Rightarrow\qquad \alpha(\textbf{p}) - \alpha(\textbf{p'}) =
  c \textbf{K}\cdot\textbf{ q},
\end{equation}
where $c$ is any real number. Then

\begin{equation}\label{}
  W_{1\alpha}(\textbf{K},\textbf{X}) =
  \int\!\!\frac{d^3q}{(2\pi)^3}\;e^{i\textbf{q}\cdot(\textbf{X} + c\textbf{K})}\rho_1(\textbf{p};\textbf{p'}).
\end{equation}
This time the shift is proportional to $\textbf{K}$ with a proportionality
coefficient which is unconstrained by momentum measurements. The corresponding
distribution in space is

\begin{equation}\label{}
  \tilde{\rho}_{1\alpha}(\textbf{x};\textbf{x}) = \int\!\!d^3K W_{1\alpha}(\textbf{K},\textbf{x})
\end{equation}
Suppose now that for $c=0$ there are no position-momentum correlations. Then
for each $\textbf{K}$ the interaction region occupies the same portion of space
and when averaged over $\textbf{K}$ coincides with that for any given
$\textbf{K}$. For $c \neq 0$, the interaction regions corresponding to
different values of $\textbf{K}$ are shifted with respect to each other and the
averaged size of the interaction region gets bigger. Its increase with respect
to the situation at $c=0$ depends on the value of $|c|$. Since this is
unconstrained by the data on momentum distributions, the radius can be made as
large as one wishes.

For instance, for the Gaussian

\begin{equation}\label{gaussp}
  \rho_1(\textbf{p};\textbf{p'}) =
  \frac{1}{(\sqrt{2\pi\Delta^2})^3}\exp\left[-\frac{\textbf{K}^2}{2\Delta^2} -
  \frac{1}{2}R^2\textbf{q}^2\right]
\end{equation}
one obtains $\tilde{\rho}_{1\alpha}(\textbf{x};\textbf{x})$ also Gaussian with
root mean square width

\begin{equation}\label{}
  R_\alpha^2 = R^2 + c^2\Delta^2.
\end{equation}
In order to see that this broadening is due to the averaging over $\textbf{K}$
it is enough to have a look at the corresponding Wigner function

\begin{equation}\label{}
  W_{1\alpha}(\textbf{K},\textbf{X}) = \frac{1}{(2\pi R\Delta)^3}\exp\left[-\frac{\textbf{K}^2}{2\Delta^2} -
  \frac{(\textbf{X}+c\textbf{K})^2}{2R^2}\right].
\end{equation}
The conclusion from this example is that without further assumptions one can at
best obtain a lower limit for the radius of the interaction region. In
practice, everybody uses, more or less consciously,  a model which supplies the
necessary additional assumptions. The problem how to choose the best model
among those which give exactly the same fit to all the data is an interesting
open problem. In order to show that this problem is not purely academic let us
consider the following example from the literature.

Some models \cite{CSZ}, \cite{BIA} assume correlations between position in
space-time and energy-momentum of the type\footnote{Sometimes called Hubble
flow}

\begin{equation}\label{hubble}
  K^\mu = \lambda X^\mu .
\end{equation}
A convenient notation is

\begin{equation}\label{}
  X_0^2 - X_\parallel^2 = \tau^2;\qquad K_0^2 - K_\parallel^2 = M_T^2\qquad
  \Rightarrow\qquad \lambda = \frac{M_T}{\tau}.
\end{equation}
Note that here $M_T$ is a temporal-longitudinal variable. Let us assume that
\cite{BIA}

\begin{equation}\label{}
  S =S_\parallel S_T
\end{equation}
where

\begin{equation}\label{}
  S_T = \exp\left[-\frac{\textbf{X}_T^2}{2r_T^2} - \frac{(\textbf{K}_T - \lambda
  \textbf{X}_T)^2}{2\delta_T^2}\right].
\end{equation}
This is a possible quantum-mechanical rendering of the transverse part of the
classical relation (\ref{hubble}). The only information about $S_\parallel$
important for our purpose is that it depends neither on $\textbf{K}_T$ nor on
$\textbf{X}_T$. The distribution of sources in the transverse plane is obtained
by integrating $S_T$ over $\textbf{K}_T$. The result is a Gaussian with
constant root mean square width $r_T$. On the other hand $S_T$ can be rewritten
in the form

\begin{equation}\label{hubcor}
  S_T = \exp\left[-\frac{\phi_T^2}{2R_D^2} - \frac{(\textbf{X}_T -
  \phi_T)^2}{2R_\phi^2}\right],
\end{equation}
where

\begin{equation}\label{}
  R_\phi = \frac{r_T}{\sqrt{1 + \mu^2}};\qquad R_D = \mu R_\phi;\qquad \mu =
  \frac{r_T}{\tau\delta_T}M_T;\qquad \phi_T = r_T\frac{\mu}{1 + \mu^2}\frac{\textbf{K}_T}{\delta_T}.
\end{equation}
This can be considered\footnote{This is an approximation, because $c$ depends
on $M_T$, it is , however, good enough for our qualitative discussion; compare
\cite{BIA}, where the full calculation can be found} as a transform with
$\alpha(\textbf{\textbf{p}}) = \frac{1}{2}c\textbf{p}^2$ and

\begin{equation}\label{}
c = \frac{r_T\mu}{\delta_T(1 + \mu^2)}
\end{equation}
of

\begin{equation}\label{nocorr}
S_{\alpha T} = \exp\left[-\frac{\textbf{K}_T^2}{2\delta_T^2}-
  \frac{\textbf{X}_T^2}{2R_\phi^2}\right].
\end{equation}
Performing the integration of $S_{\alpha T}$ over $\textbf{K}_T$ one finds for
the distribution of sources in the transverse plane a Gaussian with root mean
square width

\begin{equation}\label{}
  R_{T\alpha}^2 = R_\phi^2 = \frac{r_T^2}{1 + \mu^2},
\end{equation}
which exhibits the familiar decrease of the transverse radius with the
transverse mass as reported by so many experimental papers. Let us summarize
the situation: if our prejudice is in favor of a Hubble flow as interpreted in
\cite{BIA}, we conclude that the transverse radius does not depend on $M_T$; if
our prejudice is against correlations between position in the transverse plane
and transverse momentum, we conclude that the transverse radius decreases with
increasing $M_T$; experimental data on momentum distributions will not help us
to decide which of these two prejudices is the right one.

Our last example is one dimensional. It could e.g. apply to one transverse
component. We choose the Gaussian density matrix (\ref{gaussp}) and

\begin{equation}\label{}
  \alpha(p) = \frac{4}{3a^3}p^3\qquad \Rightarrow \qquad \alpha(p) - \alpha(p')
  = \frac{4}{3a^3}q\left(K^2 + \frac{q^3}{3}\right).
\end{equation}
A simple calculation \cite{BIZ} yields

\begin{equation}\label{}
  S_\alpha = \frac{a}{\sqrt{2\pi\Delta}}\exp\left[-\frac{K^2}{2\Delta^2} +
  B\right]Ai(A),
\end{equation}
where $Ai(\ldots)$ is the Airy function and

\begin{equation}\label{}
  A = a\tilde{X} + \frac{\omega^4}{4};\qquad B = \frac{\omega^2}{2}\left[A -
  \frac{\omega^4}{12}\right]; \qquad \omega = aR; \qquad \tilde{X} = X -
  \frac{4}{a^3}K^2.
\end{equation}
For large values of $a$ the emission function is almost Gaussian. A numerical
calculation shows that $a = 2/R$ is already large enough. For smaller values of
$a$, however, at negative $\tilde{X}$ the emission function develops big
wiggles, oscillating between positive and negative values. Its shape in the
positive $\tilde{X}$ region also significantly changes \cite{BIZ}. This example
shows how the transformations discussed in the present report can lead to
changes of the interaction region which are much more complicated than just
momentum dependent shifts.

\section{Conclusions}

The experimental data about momentum distributions tell us little about the
interaction regions, unless additional assumptions are made. The usual
recommendations: reduce the experimental errors, include more particle
correlations etc. are not enough. Exactly the same fit can be obtained from
widely different models, differing in these additional assumptions and giving
conflicting information about the interaction region. The caveat for model
users is that only some of the assumptions of the model are being tested by
comparison with the data, while others, which may be very important for drawing
inferences about the interaction region, are unconstrained by the data.

\end{document}